# Observation of anomalous Ettingshausen effect and large transverse thermoelectric conductivity in permanent magnets

Asuka Miura[1,2], Hossein Sepehri-Amin[1], Keisuke Masuda[1], Hiroki Tsuchiura[3,4], Yoshio Miura[1,5], Ryo Iguchi[1], Yuya Sakuraba[1,6], Junichiro Shiomi[2], Kazuhiro Hono[1], and Ken-ichi Uchida[1,2,4,7,a)]

[1] Research Center for Magnetic and Spintronic Materials, National Institute for Materials Science, Tsukuba 305-0047, Japan
[2] Department of Mechanical Engineering, The University of Tokyo, Tokyo 113-8656, Japan
[3] Department of Applied Physics, Tohoku University, Sendai 980-8579, Japan
[4] Center for Spintronics Research Network, Tohoku University, Sendai 980-8577, Japan
[5] Center for Spintronics Research Network, Osaka University, Osaka 560-8531, Japan
[6] PRESTO, Japan Science and Technology Agency, Saitama 332-0012, Japan
[7] Institute for Materials Research, Tohoku University, Sendai 980-8577, Japan
a)Author to whom correspondence should be addressed: UCHIDA.Kenichi@nims.go.jp

**ABSTRACT**
This study focuses on the potential of permanent magnets as thermoelectric converters. It is found that a $SmCo_5$-type magnet exhibits the large anomalous Ettingshausen effect (AEE) at room temperature and that its charge-to-heat current conversion coefficient is more than one order of magnitude greater than that of typical ferromagnetic metals. The large AEE is an exclusive feature of the $SmCo_5$-type magnet among various permanent magnets in practical use, which is independent of the conventional performance of magnets based on static magnetic properties. The experimental results show that the large AEE originates from the intrinsic transverse thermoelectric conductivity of $SmCo_5$. This finding makes a connection between permanent magnets and thermal energy engineering, providing the basis for creating "thermoelectric permanent magnets."

The Ettingshausen effect is a transverse thermoelectric conversion phenomenon in a conductor, which refers to the generation of a heat current in the direction perpendicular to a charge current and an external magnetic field.[1] With the hope of developing highly efficient thermoelectric cooling systems, the Ettingshausen effect in Bi and Bi-based compounds was investigated many decades ago.[2,3] However, using an Ettingshausen cooler is problematic as its operation requires huge external magnetic fields. One of the avenues to overcome this problem is the use of the anomalous Ettingshausen effect (AEE)[4-14] in magnetic materials, which generates a heat current in the direction of the cross product of a charge current and spontaneous magnetization:

$$\mathbf{j}_{q,AEE} = \Pi_{AEE}(\mathbf{j}_c \times \mathbf{m}), \quad (1)$$

where $\mathbf{j}_{q,AEE}$, $\mathbf{j}_c$, $\mathbf{m}$, and $\Pi_{AEE}$ denote the heat current density driven by AEE, charge current density, unit vector of the magnetization, and anomalous Ettingshausen coefficient,[11] respectively. As AEE works as a temperature modulator even in the absence of external magnetic fields and requires no complex junction structures, it may develop thermal management technologies for electronic and spintronic devices. The advantages of the AEE-based thermoelectric conversion also include the facts that the direction of heat currents can be controlled simply by changing the magnetization direction and that the thermoelectric output of the AEE device follows a simple scaling law; the cooling and heat power can be enhanced by lengthening the dimension of the device along the heat current. These advantages originate from the symmetry of Eq. (1), which were impossible with conventional thermoelectric conversion based on the Peltier effect. However, AEE has been observed only in several ferromagnetic metals.[9-14] To realize highly efficient AEE-driven thermoelectric conversion, detailed physics and material investigations using various classes of materials are necessary. Finding and developing magnetic materials with large $\Pi_{AEE}$ are also important for the realization of energy-harvesting and heat-sensing applications based on the anomalous Nernst effect (ANE),[15-23] which is the Onsager reciprocal of AEE.

In this study, we report the observation of AEE in permanent magnets at room temperature. Among the various permanent magnets in practical use, we found that the SmCo$_5$-type magnets exhibit prominently large $\Pi_{AEE}$ values, which are more than one order of magnitude larger than typical values for ferromagnetic metals, such as Ni. The figure of merit for AEE in SmCo$_5$-type magnets is comparable to the record value obtained for the ANE in a full-Heusler ferromagnet.[20] Although the thermoelectric performance needs to be improved for practical applications, our finding represents a significant step toward the realization of the AEE-based temperature modulators for the following reasons. First, in permanent magnets with large remanent magnetization and coercivity, the AEE-induced temperature change is generated in the absence of external magnetic fields in a wide temperature range (note that, in the case of SmCo$_5$-type magnets, the remanent magnetization typically remains stable up to ~300 °C). This is in sharp contrast to the conventional Ettingshausen effect, which shows good performance only at high magnetic fields and low temperatures.[2,3] Second, permanent magnets are mass-produced and widely used; adding thermoelectric conversion functionalities to permanent magnets may contribute to the development of various energy-saving technologies. Third, our results stimulate physics and material research on AEE because transverse thermoelectric phenomena in permanent magnets have not been investigated yet.

The observation of AEE in permanent magnets is realized by means of the lock-in thermography (LIT) method.[9-14,24-27] This method makes it possible to demonstrate the symmetry of AEE and quantitatively estimate its thermoelectric performance without applying magnetic fields. We measured the thermal images of the surface of permanent magnets while applying a square-wave-modulated alternating charge current $\mathbf{J}_c$ with the amplitude $J_c$, frequency $f$, and zero offset to the magnets in the $x$ direction and extracted the first harmonic response of the detected images, which are transformed into lock-in amplitude $A$ and phase $\phi$ images through Fourier analysis. Based on this procedure, the pure contribution of thermoelectric effects ($\propto J_c$) can be detected independently from the Joule-heating background ($\propto J_c^2$) [Fig. 1(f)].[9-14,26,27] To confirm the symmetry of AEE, we performed the LIT measurements for permanent magnets in both the in-plane and perpendicularly magnetized configurations at room temperature, where $\mathbf{m}$ is directed along the +$z$ and +$y$ directions, respectively [Fig. 1(a),(b)]. Following Eq. (1), the heat current driven by AEE in the in-plane (perpendicularly) magnetized configuration is generated in the $y$ ($z$) direction. The experimental details are shown in Section S1 in supplementary material.

Figure 1(c),(d) shows the $A$ and $\phi$ images of the SmCo$_5$-type magnet slab at $f$ = 2.0 Hz and $J_c$ = 0.4 A. The slab is almost fully magnetized in the absence of magnetic fields because its remanent magnetization ($\mu_0 M_r$ = 0.96 T with $\mu_0$ being the vacuum permeability) is comparable to the saturation magnetization ($\mu_0 M_s$ = 1.01 T) (Table S1 in supplementary material). In the in-plane magnetized configuration, clear current-induced temperature modulation was observed on the entire surface of the SmCo$_5$ slab [Fig. 1(c)]. As the input charge current and output temperature change oscillate with the same phase, the temperature of the sample surface increases when $\mathbf{J}_c$ flows in the −$x$ direction. Importantly, the same SmCo$_5$ slab in the perpendicularly magnetized configuration shows the temperature increase and decrease in the left and right halves, respectively [Fig. 1(d)]; the magnitude of the temperature change varies almost linearly in the $z$ direction and its sign is reversed at the center of the sample [see the $A$ and $\phi$ profiles in Fig. 1(d) and note that the $\phi$ difference between the left and right halves is ~180°]. This temperature distribution is in good agreement with the symmetry of AEE [see Eq. (1) and compare Fig. 1(c),(d) with 1(a),(b)]. We confirmed that the magnitude of the temperature modulation of the SmCo$_5$ slab is proportional to $J_c$ and consistent with the feature of AEE[9-14] [Fig. 1(e)]. As demonstrated here, LIT enables the pure detection of AEE in permanent magnets without applying external magnetic fields.

To quantitatively estimate the anomalous Ettingshausen coefficient, the behavior of the AEE-induced temperature modulation in a steady state must be known. In general, LIT images measured at low lock-in frequency show temperature distributions in nearly steady states, while those at high lock-in frequency show temperature distributions in transient states, where temperature broadening due to thermal diffusion is suppressed.[10,24] Thus, we measured the $f$ dependence of the LIT images of the SmCo$_5$ slab that gives information on the AEE-induced temperature modulation in the steady state. Figure 2 shows that the magnitude of the AEE signals in the SmCo$_5$ slab monotonically increases with decreasing $f$. The $f$ dependence of the AEE signals is well reproduced by solving the one-dimensional heat diffusion equation in the frequency domain (Section S2 in supplementary material). Based on the results in Fig. 2, the amplitude of the AEE-induced temperature modulation per unit charge current density for our SmCo$_5$ slab in the steady state, i.e., at $f = 0$ Hz, is estimated to be $A/j_c = 6.3 \times 10^{-8}$ KA$^{-1}$m$^2$, where $j_c$ is the amplitude of the square-wave-modulated charge current density. Surprisingly, this value is much greater than that of typical ferromagnetic metals; we obtained $A/j_c = 0.1 \times 10^{-8}$ KA$^{-1}$m$^2$ for a Ni slab with the same thickness and width as the SmCo$_5$ slab (see Fig. 2 and note that the magnitude of $A/j_c$ is proportional to the length along the AEE-induced heat current[13]). By substituting the steady-state AEE signals into Eq. (1), the anomalous Ettingshausen coefficient of the SmCo$_5$-type magnet is estimated to be $\Pi_\text{AEE} = 9.4 \times 10^{-4}$ V, which is more than one order of magnitude larger than that of Ni ($0.7 \times 10^{-4}$ V). The corresponding anomalous Nernst coefficient of the SmCo$_5$-type magnet at the temperature $T = 300$ K is given by $S_\text{ANE} = \Pi_\text{AEE}/T = 3.1 \times 10^{-6}$ VK$^{-1}$ via the Onsager reciprocal relation.[11]

Here, we show the relation between AEE in permanent magnets and their static magnetic properties. To clarify the dependence of AEE on the saturation magnetization, we performed the LIT measurements under the same condition using various SmCo$_5$-type magnets with different $\mu_0 M_s$ values ranging from 0.66 T to 1.01 T, where $M_s$ is reduced by partially substituting Sm with Gd and comparable to $M_r$ for all the magnets as shown in Table S1 in supplementary material (note that the SmCo$_5$ slab used for the experiments in Figs. 1 and 2 contains negligibly small amounts of Gd and the electronic structure of SmCo$_5$ near the Fermi energy is barely affected by the substitution of Sm with Gd).[28-31] Interestingly, despite the crucial difference in the magnetization and composition, we obtained almost the same $\Pi_\text{AEE}$ values for the SmCo$_5$-type magnets [Fig. 3(a)]. We also confirmed that the AEE signals are independent of the difference in the microstructure of the magnets, which affects the coercivity and resulting maximum energy product (Section S3 in supplementary material). These results indicate that the large AEE originates from an intrinsic property of SmCo$_5$. Furthermore, we found that, among the three types of rare-earth permanent magnets in practical use, i.e., the SmCo$_5$-, Sm$_2$Co$_{17}$-, and Nd$_2$Fe$_{14}$B-type magnets, only the SmCo$_5$-type magnets exhibit the large AEE, although the Sm$_2$Co$_{17}$- and Nd$_2$Fe$_{14}$B-type magnets have larger magnetization [Fig. 3(a)]. This behavior clearly deviates from the scaling between the magnetization and transverse thermoelectric effects.[18]

Significantly, the performance of the SmCo$_5$-type magnets as the AEE/ANE-based thermoelectric converters reaches a record high. The figure of merit for AEE/ANE is defined as[11,32]

$$Z_\text{AEE}T = \frac{\Pi_\text{AEE}^2 \sigma_{xx}}{\kappa} \frac{1}{T} \quad \left(= \frac{S_\text{ANE}^2 \sigma_{xx}}{\kappa} T\right). \tag{2}$$

By measuring the longitudinal electric conductivity $\sigma_{xx}$ and the thermal conductivity $\kappa$, we estimated the $Z_\text{AEE}T$ values of the magnets. As shown in Fig. 3(b), we found no correlation between $Z_\text{AEE}T$ and $\mu_0 M_s$ in each type of magnet. The maximum $Z_\text{AEE}T$ value of the SmCo$_5$-type magnet is $4.5 \times 10^{-4}$ at $T = 300$ K, which is two orders (one order) of magnitude greater than that of Ni (Sm$_2$Co$_{17}$- and Nd$_2$Fe$_{14}$B-type magnets) and comparable to the value of full-Heusler Co$_2$MnGa that exhibits the remarkably large ANE.[20-22]

Now, we are in a position to discuss the origin of the large AEE in the SmCo$_5$-type magnets. Based on the Onsager reciprocal relation between AEE and ANE, the anomalous Ettingshausen coefficient can be divided into the following two terms:[20,23]

$$\Pi_\text{AEE} = \left(\rho_{xx}\alpha_{xy} + \rho_{xy}\alpha_{xx}\right)T \equiv \Pi_\text{I} + \Pi_\text{II}, \tag{3}$$

where $\rho_{xx} = 1/\sigma_{xx}$ ($\rho_{xy} = -\sigma_{xy}/\sigma_{xx}^2$) denotes the diagonal (off-diagonal) component of the electric resistivity tensor, $\alpha_{xx}$ ($\alpha_{xy}$) the diagonal (off-diagonal) component of the thermoelectric conductivity tensor, $\Pi_\mathrm{I} = \rho_{xx}\alpha_{xy}T$, and $\Pi_\mathrm{II} = \rho_{xy}\alpha_{xx}T$. In Eq. (3), we assume a fully magnetized ferromagnet and disregard the magnetic-field-dependent contributions in the electric and thermoelectric transport coefficients. The contribution of $\Pi_\mathrm{I}$ is often regarded as an intrinsic part of AEE as it originates from the transverse thermoelectric conductivity $\alpha_{xy}$, which is determined by the energy derivative of the anomalous Hall conductivity $\sigma_{xy}$ (Section S4 in supplementary material).[20,23] In contrast, the contribution of $\Pi_\mathrm{II}$ is attributed to the concerted action of the Seebeck effect and anomalous Hall effect (AHE), and can be rewritten as $\Pi_\mathrm{II} = S_{xx}T\tan\theta_\mathrm{AHE}$, where $S_{xx} = \rho_{xx}\alpha_{xx}$ and $\theta_\mathrm{AHE} = \rho_{xy}/\rho_{xx}$ are the Seebeck coefficient and anomalous Hall angle, respectively. To estimate $\Pi_\mathrm{II}$, we measured the Seebeck effect and AHE in the $SmCo_5$-, $Sm_2Co_{17}$-, and $Nd_2Fe_{14}B$-type magnets at room temperature. Figure 4(a) shows the Hall resistivity $\rho_\mathrm{Hall}$ of the magnets as a function of the magnetic field $\mu_0H$. The magnets exhibit clear hysteresis loops in the $\mu_0H$-$\rho_\mathrm{Hall}$ curves; the AHE contribution can be extracted by extrapolating the $\rho_\mathrm{Hall}$ data in the high-field region to the zero field. The obtained $\theta_\mathrm{AHE}$ values are negative for all the magnets and the magnitude of $\theta_\mathrm{AHE}$ of the $SmCo_5$-type magnet is smaller than that of the $Sm_2Co_{17}$- and $Nd_2Fe_{14}B$-type magnets [Fig. 4(b)]. The inset to Fig. 4(b) shows that $\sigma_{xy}$ of the magnets is almost independent of $\sigma_{xx}$, suggesting that AHE in these materials is in the intrinsic regime in the scaling relation.[15] By combining the AHE data with the Seebeck coefficient in Fig. 4(c), we obtained the $\Pi_\mathrm{II}$ values for the $SmCo_5$-, $Sm_2Co_{17}$-, and $Nd_2Fe_{14}B$-type magnets at $T = 300$ K. As shown in Fig. 4(d), $\Pi_\mathrm{II}$ is positive for the magnets, the sign of which is the same as (opposite to) that of $\Pi_\mathrm{AEE}$ for the $SmCo_5$- and $Sm_2Co_{17}$-type magnets ($Nd_2Fe_{14}B$-type magnet). The magnitude of $\Pi_\mathrm{II}$ for the $SmCo_5$-type magnet is much smaller than that of $\Pi_\mathrm{AEE}$, indicating that AEE is dominated by the $\Pi_\mathrm{I}$ term owing to the large transverse thermoelectric conductivity [Fig. 4(d)]. The estimated $\alpha_{xy}$ value of the $SmCo_5$-type magnet (4.6 $K^{-1}Am^{-1}$) is much larger than those of the $Sm_2Co_{17}$-type magnet (0.6 $K^{-1}Am^{-1}$) and the $Nd_2Fe_{14}B$-type magnet (–0.9 $K^{-1}Am^{-1}$) and, surprisingly, even larger than that of $Co_2MnGa$ at room temperature (2.4-3.0 $K^{-1}Am^{-1}$).[20] This remarkably large $\alpha_{xy}$ can be the intrinsic property of $SmCo_5$, which is supported by our first-principles calculations; the $\alpha_{xy}$ value calculated from the electronic structures of $SmCo_5$ is comparable to the experimental value (Section S4 in supplementary material). However, more detailed theoretical investigations, such as the further analyses of the Berry curvature and its contribution to AEE, are necessary to clarify the microscopic mechanism of the transverse thermoelectric response in the $SmCo_5$-type magnets.

Although the $SmCo_5$-type magnet exhibits a record-high $\alpha_{xy}$ value, its thermoelectric performance is still insufficient for practical applications, and further investigations are required to explore and develop magnets with larger $Z_\mathrm{AEE}T$. Judging from the previous discussions on the thermoelectric conversion efficiency for ANE, $Z_\mathrm{AEE}T > 0.1$ is necessary for practical applications.[32,33] Based on the results for $SmCo_5$, $Z_\mathrm{AEE}T > 0.1$ can be obtained, for example, by realizing 10 times enhancement of $\Pi_\mathrm{AEE}$ and 50% reduction of $\kappa$ simultaneously [see Eq. (2)]. A clue for the improvement of $\Pi_\mathrm{AEE}$ is included in our first-principles calculations; carrier doping of $SmCo_5$ may modulate the transverse transport coefficients and increase the resulting $\Pi_\mathrm{AEE}$ (Section S4 in supplementary material). Furthermore, as AEE originates from the spin-orbit interaction acting on spin-polarized electron transport, the substitution of Co with heavy elements possessing strong spin-orbit interaction, such as Pt and Pd, can be another strategy for enhancing the intrinsic contribution of $\Pi_\mathrm{AEE}$. The $\kappa$ reduction is achievable by nanostructuring[34] and grain boundary engineering,[35] although the $SmCo_5$-type magnet used in this study has no grain boundary phases (Section S3 in supplementary material). The grain boundary engineering may also modulate extrinsic spin-orbit interaction and spin-transport properties in the magnets,[35] which have a potential to improve the transverse thermoelectric response.

In conclusion, we investigated the transverse thermoelectric conversion properties in the well-known rare-earth magnets and observed the large AEE in the $SmCo_5$-type magnets. This result is the first step towards the creation of "thermoelectric permanent magnets"; such multi-functional materials will invigorate fundamental studies in spin caloritronics and thermoelectrics, and develop thermal energy management and harvesting technologies for electronic and spintronic devices. Importantly, in contrast to conventional soft magnetic materials, thermoelectric permanent magnets can modulate temperature or generate electricity in the absence of external magnetic fields. Although the selection of conventional magnets is determined by the maximum energy product and maximum usable temperature, this multi-functionality and versatility may change the selection criteria for permanent magnets and lead to unconventional applications. We thus anticipate that the proof-of-concept demonstration reported here will provide a basis for exploring and developing magnets with larger $Z_{AEE}T$.

The authors thank Y. Hirayama, X. Tang, T. Seki, M. Matsumoto, H. Adachi, M. Shimizu, and H. O. Jeschke for valuable discussions. This work was supported by CREST "Creation of Innovative Core Technologies for Nano-enabled Thermal Management" (JPMJCR17I1), PRESTO "Phase Interfaces for Highly Efficient Energy Utilization" (JPMJPR12C1), and PRESTO "Scientific Innovation for Energy Harvesting Technology" (JPMJPR17R5) from JST, Japan; Grant-in-Aid for Scientific Research (S) (JP18H05246) and Grant-in-Aid for Early-Career Scientists (JP18K14116) from JSPS KAKENHI, Japan; and the NEC Corporation.  A.M. is supported by JSPS through Research Fellowship for Young Scientists (JP18J02115).

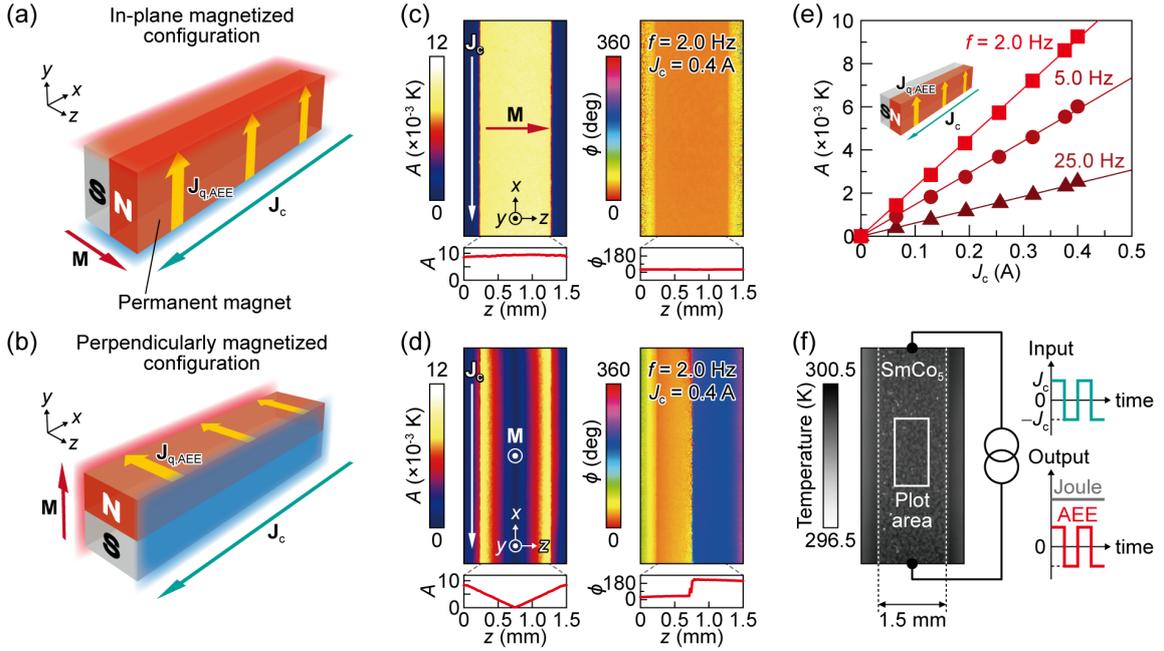

**FIG. 1.** (a),(b) Schematic illustrations of AEE in a permanent magnet in the in-plane magnetized (a) and perpendicularly magnetized (b) configurations. $J_c$ and $J_{q,AEE}$ denote the charge current applied to the magnet and heat current driven by AEE, respectively. In both the configurations, **M** shows the direction of the remanent magnetization, which is along the $c$ axis of the magnet. (c),(d) Lock-in amplitude $A$ and phase $\phi$ images for the SmCo$_5$ slab in the in-plane magnetized (c) and perpendicularly magnetized (d) configurations at the charge current magnitude $J_c = 0.4$ A and lock-in frequency $f = 2.0$ Hz. The surface $A$ and $\phi$ profiles along the $z$ direction are also shown, which are extracted around the center of the thermal images. (e) $J_c$ dependence of $A$ for the SmCo$_5$ slab in the in-plane magnetized configuration at $f = 2.0$, $5.0$, and $25.0$ Hz. (f) Steady-state infrared image for the black-ink-coated SmCo$_5$ slab and setup for the LIT measurements. The data points in (e) were obtained by averaging the temperature modulation signals on the area defined by the white rectangle with the size of $50 \times 100$ pixels in (f). The results shown in this figure were obtained from the SmCo$_5$ slab with the saturation magnetization of $\mu_0 M_s = 1.01$ T.

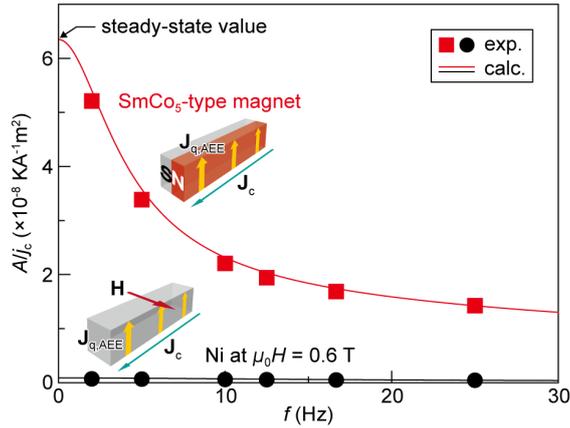

**FIG. 2.** $f$ dependence of $A/j_c$ for the SmCo$_5$ slab with $\mu_0 M_s = 1.01$ T (red squares) and the Ni slab (black circles) in the in-plane magnetized configuration. $j_c$ is the amplitude of the square-wave-modulated charge current density applied to the slabs, which is smaller than the sinusoidal amplitude by a factor of $\pi/4$ (note that the sinusoidal amplitude of the charge current density was used to estimate $\Pi_{AEE}$ from the LIT data). The LIT measurements for the SmCo$_5$ slab were performed in the absence of the external magnetic field **H**, while the data for the Ni slab were measured with applying **H** with the magnitude $\mu_0 H = 0.6$ T in the direction perpendicular to $J_c$. At 0.6 T, the magnetization of the Ni slab aligns along the **H** direction. The solid red and black lines show the calculated $f$ dependence of $A/j_c$ for the SmCo$_5$ and Ni slabs, respectively, which were obtained by solving the one-dimensional heat diffusion equation.

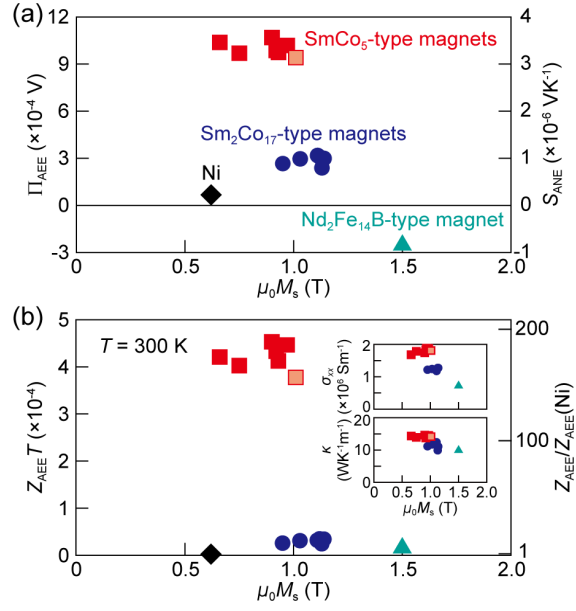

**FIG. 3.** (a) $\mu_0 M_s$ dependence of the anomalous Ettingshausen coefficient $\Pi_{AEE}$ and corresponding anomalous Nernst coefficient $S_{ANE}$ (= $\Pi_{AEE}/T$) for the SmCo$_5$-type magnets (red squares), Sm$_2$Co$_{17}$-type magnets (blue circles), Nd$_2$Fe$_{14}$B-type magnet (green triangle), and Ni (black diamond) at temperature $T$ = 300 K. The data point for the SmCo$_5$-type magnet with $\mu_0 M_s$ = 1.01 T, used for the experiments in Figs. 1 and 2, is emphasized by changing its color. All the AEE measurements shown in this figure were performed in the in-plane magnetized configuration. (b) $\mu_0 M_s$ dependence of the dimension-less figure of merit $Z_{AEE}T$ for AEE in the magnets at $T$ = 300 K and the $Z_{AEE}$ values normalized by the value of Ni. The insets to (b) show the $\mu_0 M_s$ dependence of the longitudinal electric conductivity $\sigma_{xx}$ and the thermal conductivity $\kappa$ of the magnets. Despite being out of range in the insets, the $\sigma_{xx}$ and $\kappa$ values of Ni were observed to be $12.2 \times 10^6$ Sm$^{-1}$ and 76.4 WK$^{-1}$m$^{-1}$, respectively.
88

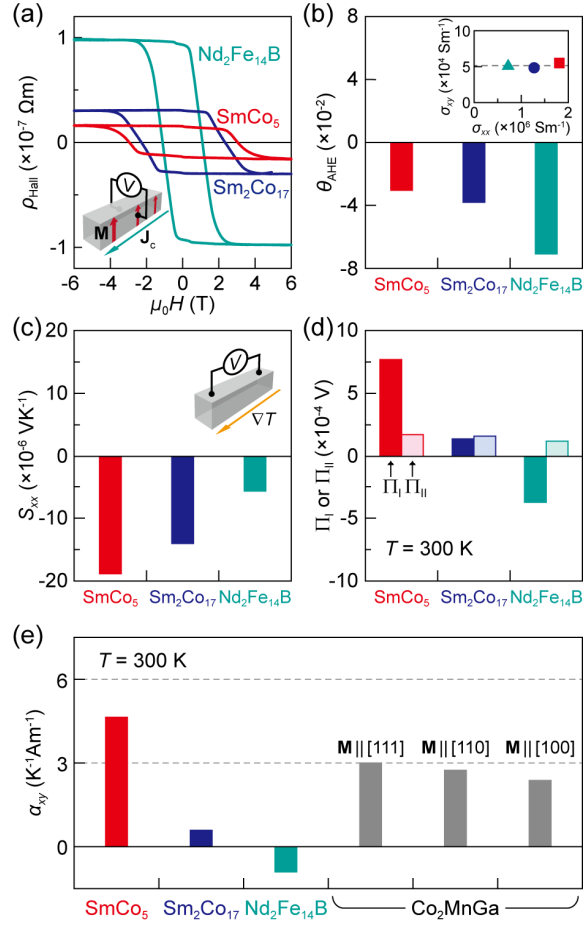

**FIG. 4.** (a) $\mu_0 H$ dependence of the Hall resistivity $\rho_{Hall}$ for the $SmCo_5$-, $Sm_2Co_{17}$-, and $Nd_2Fe_{14}B$-type magnets. The results shown in this figure were obtained from the $SmCo_5$-type, $Sm_2Co_{17}$-, and $Nd_2Fe_{14}B$-type magnets with $\mu_0 M_s$ = 1.01 T, 1.14 T, and 1.50 T, respectively (Table S1 in supplementary material). (b) Anomalous Hall angle $\theta_{AHE}$ of the magnets. The inset to (b) shows the $\sigma_{xx}$ dependence of the transverse electric conductivity $\sigma_{xy}$ for the $SmCo_5$-type magnet (red square), $Sm_2Co_{17}$-type magnet (blue circle), and $Nd_2Fe_{14}B$-type magnet (green triangle), where the $\sigma_{xy}$ values were obtained by extrapolating the $\rho_{Hall}$ data in the high-field region (9 T < $\mu_0 H$ < 10 T) to the zero field. (c) Seebeck coefficient $S_{xx}$ of the magnets. The $S_{xx}$ values were estimated by measuring the electric voltage $V$ along the temperature gradient $\nabla T$. We confirmed that the magneto-Seebeck effect[8,10] in the magnets is negligibly small by repeating the thermopower measurements after demagnetization. (d) $\Pi_{I}$ and $\Pi_{II}$, defined by Eq. (3), of the magnets at $T$ = 300 K. (e) Transverse thermoelectric conductivity $\alpha_{xy}$ at $T$ = 300 K for the $SmCo_5$-, $Sm_2Co_{17}$-, and $Nd_2Fe_{14}B$-type magnets and the full-Heusler $Co_2MnGa$. The $\alpha_{xy}$ values of $Co_2MnGa$ for different configurations were estimated from Fig. 3 of Ref. 20, where **M** is parallel to the [111], [110], or [100] direction of $Co_2MnGa$.

# Supplementary Material

**S1. Experimental details**

The polycrystalline SmCo$_5$- and Sm$_2$Co$_{17}$-type magnet slabs used in this study are commercially available from Magfine Corporation, Japan. The magnetic properties and composition of the SmCo$_5$- and Sm$_2$Co$_{17}$-type magnets are shown in Table S1. The polycrystalline Ni slab is commercially available from The Nilaco Corporation, Japan. The Nd$_2$Fe$_{14}$B-type magnet slab is a commercially-available-class anisotropic sintered magnet prepared by conventional powder processing using a jet milling process, followed by magnetic field alignment of the powders, liquid sintering at 1080 °C, and post annealing at 600 °C for 1 h. The chemical composition of the anisotropic Nd$_2$Fe$_{14}$B-type sintered magnet is Nd$_{11.7}$Pr$_{2.8}$Fe$_{75.8}$B$_{6.0}$Co$_{1.0}$Al$_{0.5}$Cu$_{0.1}$O$_{2.1}$ (at%), measured using inductively coupled plasma analysis. The experimental results shown in the main text were obtained by using the samples having a rectangular cuboid shape; the lengths of the SmCo$_5$-, Sm$_2$Co$_{17}$-, and Nd$_2$Fe$_{14}$B-type magnets (Ni slab) along the $x$, $y$, and $z$ directions are 12.0 mm (20.0 mm), 1.5 mm (1.5 mm), and 1.5 mm (1.5 mm), respectively. During the LIT measurements, the samples were fixed on a plastic plate with low thermal conductivity to reduce the heat loss due to thermal conduction as much as possible. To enhance the infrared emissivity and ensure uniform emission properties, the top surface of the samples was coated with insulating black ink with an emissivity of > 0.95.

The longitudinal electric conductivity of the samples was measured by the four probe method. The anomalous Hall effect was measured with the physical property measurement system (PPMS, Quantum Design, Inc.). The Seebeck coefficient was measured with the Seebeck coefficient/electric resistance measurement system (ZEM-3, ADVANCE RIKO, Inc.). These measurements were performed at room temperature by using the rectangular-shaped samples with the same size as those used for the measurements of the anomalous Ettingshausen effect (AEE). The thermal conductivity of the permanent magnets at room temperature was estimated through thermal diffusivity measurements using the laser flash method and specific heat measurements using the differential scanning calorimetry. Here, the thermal diffusivity in the direction perpendicular to the $c$ axis was measured by using the rectangular-shaped samples with a size of 10.0 × 10.0 × 2.0 mm$^3$, where the $c$ axis is along the 10.0-mm direction.

**Table S1.** Magnetic properties and composition of permanent magnets. The element contents of the SmCo$_5$- and Sm$_2$Co$_{17}$-type magnets with various $\mu_0 M_s$ values were measured by inductively coupled plasma atomic emission spectroscopy. $M_r$, $H_{cb}$, and $(BH)_{max}$ denote the remanent magnetization, coercivity, and maximum energy product, respectively. The magnetic properties were measured by vibrating sample magnetometry.

| Magnet type | $\mu_0 M_s$ (T) | $\mu_0 M_r$ (T) | $\mu_0 H_{cb}$ (T) | $(BH)_{max}$ (kJm$^{-3}$) | Content (at%) | | | | | | | | |
|---|---|---|---|---|---|---|---|---|---|---|---|---|---|
| | | | | | Sm | Co | Fe | Gd | Cu | Zr | Pr | Sn | O |
| SmCo$_5$ | 0.66 | 0.62 | 0.61 | 76 | 8.3 | 82.3 | 0.4 | 8.8 | 0.0 | 0.0 | 0.0 | 0.3 | 3.4 |
| | 0.75 | 0.71 | 0.70 | 99 | 11.0 | 82.1 | 0.3 | 6.3 | 0.0 | 0.0 | 0.0 | 0.3 | 2.3 |
| | 0.92 | 0.89 | 0.71 | 135 | 14.3 | 82.1 | 0.3 | 3.1 | 0.0 | 0.0 | 0.0 | 0.2 | 1.7 |
| | 0.90 | 0.87 | 0.83 | 147 | 14.1 | 82.2 | 0.3 | 3.1 | 0.0 | 0.0 | 0.0 | 0.2 | 2.1 |
| | 0.93 | 0.90 | 0.84 | 157 | 15.4 | 82.1 | 0.3 | 2.1 | 0.0 | 0.0 | 0.0 | 0.2 | 2.0 |
| | 0.97 | 0.94 | 0.88 | 167 | 15.4 | 82.1 | 0.3 | 2.1 | 0.0 | 0.0 | 0.0 | 0.2 | 1.9 |
| | 1.01 | 0.96 | 0.78 | 163 | 17.2 | 82.1 | 0.3 | 0.2 | 0.0 | 0.0 | 0.0 | 0.2 | 2.1 |
| Sm$_2$Co$_{17}$ | 0.95 | 0.89 | 0.85 | 150 | 10.0 | 59.1 | 18.6 | 2.1 | 6.6 | 2.2 | 1.4 | 0.0 | 2.5 |
| | 1.11 | 1.06 | 1.00 | 211 | 11.9 | 59.3 | 18.8 | 0.0 | 6.4 | 2.2 | 1.4 | 0.0 | 1.7 |
| | 1.14 | 1.09 | 1.01 | 221 | 11.8 | 59.6 | 19.0 | 0.0 | 5.9 | 2.2 | 1.4 | 0.0 | 1.5 |
| | 1.03 | 0.99 | 0.94 | 186 | 12.0 | 60.5 | 19.2 | 0.0 | 6.0 | 2.2 | 0.0 | 0.0 | 1.6 |
| | 1.13 | 1.10 | 0.98 | 232 | 11.6 | 59.7 | 20.8 | 0.0 | 5.5 | 2.4 | 0.0 | 0.0 | 1.5 |
| | 1.12 | 1.08 | 0.65 | 212 | 12.0 | 60.6 | 19.2 | 0.0 | 6.0 | 2.2 | 0.0 | 0.0 | 1.6 |

## S2. Calculations of lock-in frequency dependence

To estimate the AEE signals in the steady-state condition, we calculated the $f$ dependence of the AEE-induced temperature modulation by solving the one-dimensional heat diffusion equation in the frequency domain. The model system used for the calculation is an isolated bulk ferromagnet in which only the length $L$ (1.5 mm) along the AEE-induced heat current density is considered. As the boundary condition, the total heat current at the ends of the ferromagnet is set to be zero. The temperature modulation at one end of the ferromagnet can be compared with the experimental results, because the samples used for the LIT measurements are thermally isolated. Here, the temperature modulation is given by

$$Ae^{-i\phi} = \sqrt{\frac{i}{2\pi f C \rho \kappa}} \frac{\cos\left(-L\sqrt{2\pi f C \rho / i\kappa}\right) - 1}{\sin\left(-L\sqrt{2\pi f C \rho / i\kappa}\right)} \tilde{j}_{q,\text{AEE}}, \quad (S1)$$

where $C$, $\rho$, and $\tilde{j}_{q,\text{AEE}}$ are the specific heat, density, and first-harmonic sinusoidal amplitude of the AEE-induced heat current density, respectively. To calculate the $f$ dependence of the temperature modulation, the measured values of $C$, $\rho$, and $\kappa$ were substituted into Eq. (S1). The calculation results are in good agreement with the observed $f$ dependence of the AEE signals (Fig. 2 in the main text).

## S3. Microstructure analysis of SmCo$_5$-type magnets

We observed the microstructure of two SmCo$_5$-type sintered magnets with different $\mu_0 M_s$ values of 0.66 T and 1.01 T, of which the composition is Sm$_{8.3}$Gd$_{8.8}$Co$_{82.3}$Fe$_{0.4}$Sn$_{0.3}$O$_{3.4}$ and Sm$_{17.2}$Gd$_{0.2}$Co$_{82.1}$Fe$_{0.3}$Sn$_{0.2}$O$_{2.1}$ (at%), respectively (Table S1). Large Gd content in the SmCo$_5$-type magnet reduces the magnetization and enhances the magnetic anisotropy field, responsible for the decrease of the maximum energy product.[28,29] The back-scattered electron (BSE)-scanning electron microscopy (SEM) images and corresponding SEM-energy dispersive spectroscopy (EDS) maps for the SmCo$_5$-type magnets are shown in Fig. S1. Both the samples contain (Sm + Gd)-rich or Sm-rich grains in the form of oxide (note that the SEM-EDS map of O is not shown). The Sm + Gd and Co EDS maps of the small $M_s$ sample clarifies a uniform distribution of Sm and Gd in the matrix that corresponds to the (SmGd)Co$_5$ phase [Fig. S1(b)]. In contrast, a secondary phase with a bright gray contrast appears in the BSE-SEM image of the large $M_s$ sample [Fig. S1(e)]. The EDS line profile from this phase is plotted in Fig. S1(i) showing the existence of a Sm-rich phase in the microstructure of the large $M_s$ sample. The high resolution high angle annular dark field (HAADF)-scanning transmission electron microscopy (STEM) images in Fig. S1(d),(h) show a direct contact of two (SmGd)Co$_5$ or SmCo$_5$ grains without existence of any grain boundary phases for both the samples. Figure S1(k) shows the HAADF-STEM image obtained from an interface of Sm-rich secondary phase with the SmCo$_5$ grain of the large $M_s$ sample. This result also shows a direct contact between two neighboring grains without formation of any grain boundary phases. This situation is different from the microstructure of the Nd$_2$Fe$_{14}$B-type sintered magnets.[S1,S2] The projection of Sm atoms in Fig. S1(k), further clarified with atomic resolution STEM-EDS in Fig. S1(j), shows the existence of numerous stacking faults in the basal plane of the Sm-rich phase in the large $M_s$ sample. The selected area electron diffraction pattern shown in the inset to Fig. S1(k) clarifies that the Sm-rich phase in the large $M_s$ sample corresponds to the Sm$_5$Co$_{19}$ phase. Even though different phases were found in the microstructure of the SmCo$_5$-type magnets with different saturation magnetization and total Gd content, the magnets exhibit comparable AEE signals (Fig. 3 in the main text), indicating that the observed large AEE does not have microstructure origin.

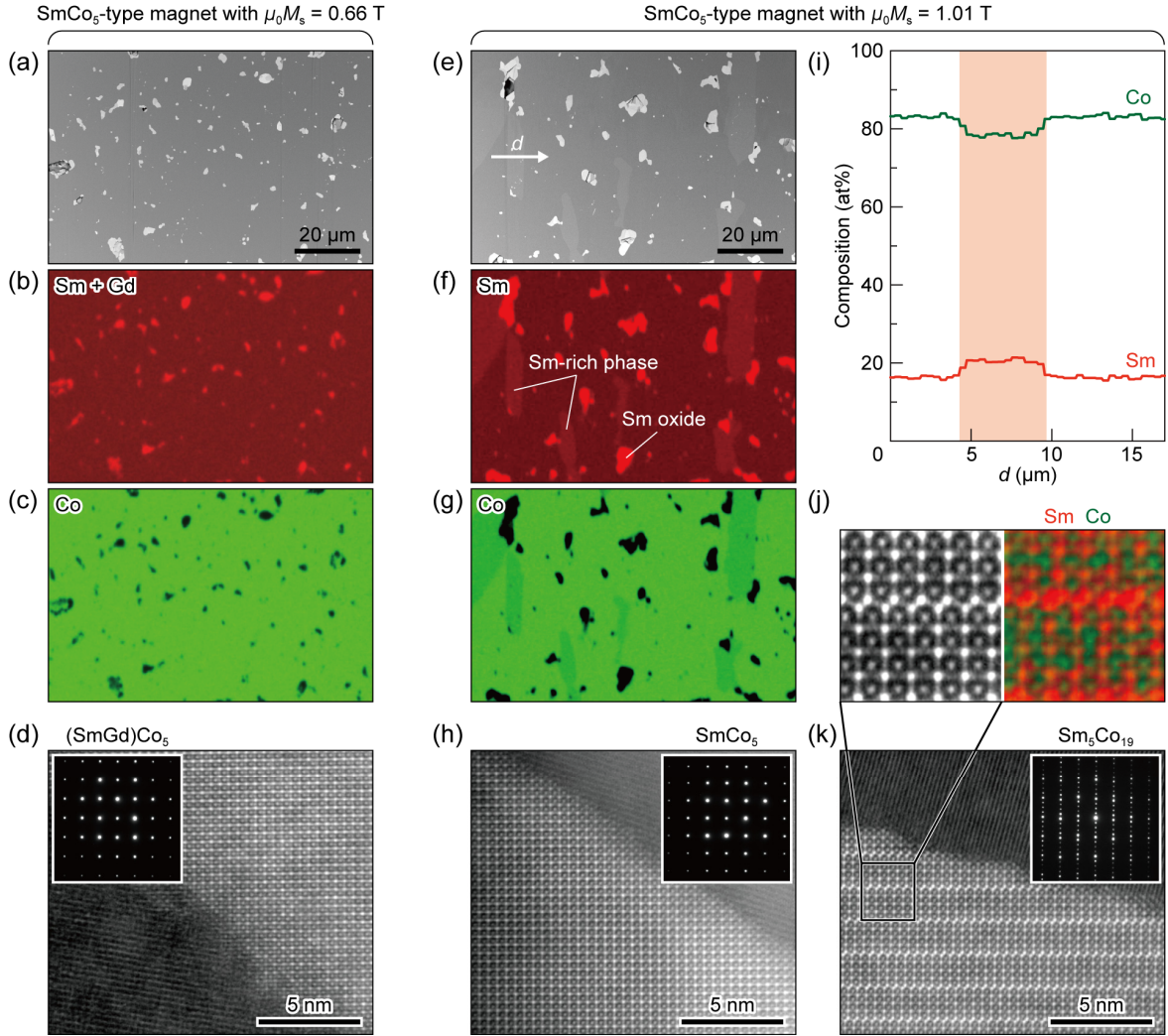

**FIG. S1.** (a)-(c) BSE-SEM image (a) and corresponding SEM-EDS maps of Sm + Gd (b) and Co (c) for the SmCo$_5$-type magnet with the saturation magnetization of $\mu_0 M_s$ = 0.66 T. (d) High resolution HAADF-STEM image showing grain boundaries for the SmCo$_5$-type magnet with $\mu_0 M_s$ = 0.66 T. (e)-(g) BSE-SEM image (e) and corresponding SEM-EDS maps of Sm (f) and Co (g) for the SmCo$_5$-type magnet with $\mu_0 M_s$ = 1.01 T. As shown in Table S1, the composition of the SmCo$_5$-type magnet with $\mu_0 M_s$ = 1.01 T is close to pure SmCo$_5$. (h) HAADF-STEM image for the SmCo$_5$ phase in the SmCo$_5$-type magnet with $\mu_0 M_s$ = 1.01 T. (i) EDS line profiles measured from the arrow "$d$" shown in (e). (j) Higher magnification HAADF-STEM image and corresponding STEM-EDS map of Sm and Co for the Sm$_5$Co$_{19}$ phase in the SmCo$_5$-type magnet with $\mu_0 M_s$ = 1.01 T. (k) HAADF-STEM image showing grain boundaries for the Sm$_5$Co$_{19}$ phase in the SmCo$_5$-type magnet with $\mu_0 M_s$ = 1.01 T. The selected area electron diffraction patterns shown in the insets to the HAADF-STEM images were obtained for the (SmGd)Co$_5$ or SmCo$_5$ phase along [110] zone axis (d),(h) and for the Sm$_5$Co$_{19}$ phase (k).

## S4. First-principles calculations of transverse transport coefficients

To support our interpretation, the transverse transport coefficients of SmCo$_5$ were investigated by means of the first-principles calculations combined with the linear response theory. We first calculated the electronic structures of SmCo$_5$ by means of the full-potential linearized augmented plane wave method including the effect of the spin-orbit interaction, which is implemented in the WIEN2k program.[S3] The lattice constants of the primitive unit cell were fixed to $a$ = 5.01 Å and $c$ = 3.97 Å [Fig. S2(a)]. In the calculation, we took into account the Coulomb interaction $U$ and the Hund coupling $J$ for $f$ orbitals in Sm, since the strong electron correlation is one of the most important features in $f$-electron systems.[S4,S5] Here, we chose $U$ = 9.00 eV in order to obtain a reasonable orbital magnetic moment of Sm[S6] and set $J$ = 0.75 eV after confirming the weak $J$ dependence of our results.

The self-consistent-field calculation using 19 × 19 × 21 $k$ points gave an antiferromagnetic state with the antiparallel alignment of the spin magnetic moments between Sm and Co atoms, which is energetically favored compared to the ferromagnetic state.[S7] We calculated the anomalous Hall conductivity $\sigma_{xy}$ using the following expression derived from the Kubo formula:[S8,S9]

$$\sigma_{xy}(\varepsilon) = \frac{e^2\hbar}{m^2}\int\frac{d^3k}{(2\pi)^3}\sum_{n\neq n'}\left[\theta(E_{n',\mathbf{k}},\varepsilon) - \theta(E_{n,\mathbf{k}},\varepsilon)\right]\frac{\operatorname{Im}\langle\psi_{n,\mathbf{k}}|p_x|\psi_{n',\mathbf{k}}\rangle\langle\psi_{n',\mathbf{k}}|p_y|\psi_{n,\mathbf{k}}\rangle}{\left(E_{n',\mathbf{k}} - E_{n,\mathbf{k}}\right)^2}, \quad (S2)$$

where $p_x$ ($p_y$) is the $x$ ($y$) component of the momentum operator, $n$ ($n'$) is the band index for the occupied (unoccupied) state, $\psi_{n,\mathbf{k}}$ is the eigenstate with the eigenenergy $E_{n,\mathbf{k}}$, and $\theta(E_{n,\mathbf{k}}, \varepsilon)$ is the occupation function for the band $n$ and the wave vector $\mathbf{k}$ at the energy $\varepsilon$ relative to the Fermi energy. We used 67 × 67 × 74 $k$ points for the Brillouin-zone integration ensuring good convergence for $\sigma_{xy}$. The transverse thermoelectric conductivity $\alpha_{xy}$ for a given temperature $T$ was calculated by using the following expression:

$$\alpha_{xy} = -\frac{1}{eT}\int d\varepsilon\left(-\frac{\partial f}{\partial \varepsilon}\right)(\varepsilon - \mu)\sigma_{xy}(\varepsilon), \quad (S3)$$

where $f = 1/[\exp((\varepsilon - \mu)/k_BT) + 1]$ is the Fermi distribution function. In the calculations of $\sigma_{xy}$ and $\alpha_{xy}$, the **M** direction is set to be along the $c$ axis, which is the magnetic easy axis of SmCo$_5$ [Fig. S2(a)].[28] The $y$ direction in Eqs. (S2) and (S3) corresponds to the $b$ axis in Fig. S2(a). Here, the $\sigma_{xy}$ and $\alpha_{xy}$ values at the Fermi energy, $\mu = 0$ eV, can be compared with the experimental results.

Figure S2(b),(c) shows the calculated $\sigma_{xy}$ and $\alpha_{xy}$ values at $T$ = 300 K as a function of the chemical potential $\mu$, obtained from the electronic structures of SmCo$_5$. The $\sigma_{xy}$ and $\alpha_{xy}$ values at $\mu$ = 0 eV are estimated to be 5.3 × 10$^4$ Sm$^{-1}$ and 6.4 K$^{-1}$Am$^{-1}$, respectively, which are comparable to the experimental values of $\sigma_{xy}$ = 5.5 × 10$^4$ Sm$^{-1}$ and $\alpha_{xy}$ = 4.6 K$^{-1}$Am$^{-1}$ obtained for the SmCo$_5$-type magnet with $\mu_0M_s$ = 1.01 T [compare Fig. S2(b),(c) with the insets to Fig. 4(b),(d)]. We also found that the calculated $\alpha_{xy}$ value for SmCo$_5$ is almost the same as that for GdCo$_5$ because the electronic structure of these materials near the Fermi energy is determined by the Co sublattice [Fig. S2(c)], which can explain the experimental fact that AEE in the SmCo$_5$-type magnets is little affected by the substitution of Sm with Gd (Fig. 3). The consistency between the calculations and experiments confirms the scenario that the large AEE observed in this study is the intrinsic property of SmCo$_5$.

Importantly, a clue for further improvement of AEE is included in our first-principles calculations. The strong $\mu$ dependence of $\sigma_{xy}$ and $\alpha_{xy}$ indicates that carrier doping of SmCo$_5$ may modulate these transverse transport coefficients and increase the resulting $\Pi_{AEE}$ [see Fig. S2(b),(c) and note that we obtained the unprecedentedly large transverse thermoelectric conductivity of $\alpha_{xy}$ = 9.1 K$^{-1}$Am$^{-1}$ at the slightly modulated chemical potential of $\mu$ = 0.04 eV].

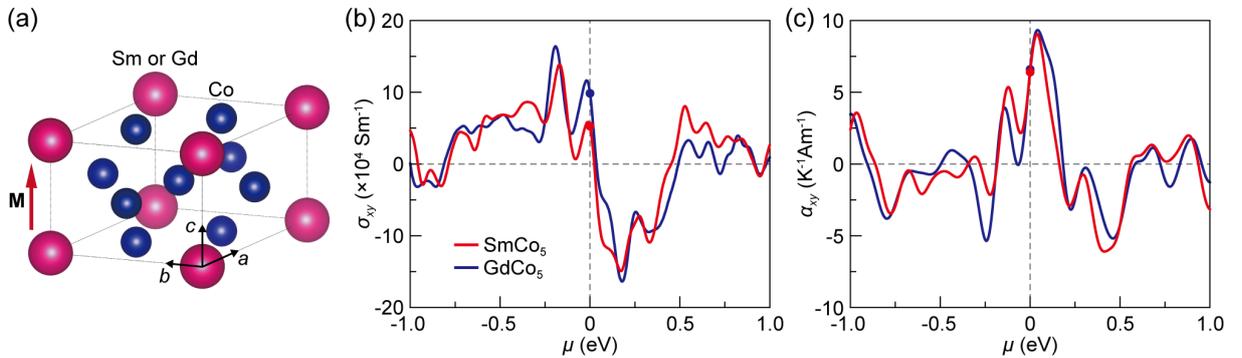

**FIG. S2.** (a) Crystal structure of SmCo$_5$ or GdCo$_5$. **M** shows the direction of the net magnetization in the calculations. (b) Calculated chemical potential $\mu$ dependence of $\sigma_{xy}$ for SmCo$_5$ (red curve) and GdCo$_5$ (blue curve). (c) Calculated $\mu$ dependence of $\alpha_{xy}$ for SmCo$_5$ and GdCo$_5$.